\documentclass[]{llncs}

\usepackage[utf8]{inputenc}      
\usepackage[T1]{fontenc}         
\usepackage{microtype}           
\usepackage[english]{babel}      
\usepackage{graphicx}            
\usepackage{cite}                
\usepackage[]{hyperref}            
\usepackage{comment}             
\usepackage{newtxtext,newtxmath} 

\title{Event Stream Processing with Multiple Threads\thanks{This paper is an extended version of the paper of same title published at the 2017 International Conference on Runtime Verification.}}

\author{Sylvain Hallé \and Raphaël Khoury \and Sébastien Gaboury%
}
\institute{%
Laboratoire d'informatique formelle \\
Université du Québec à Chicoutimi, Canada%
}

\usepackage{amsmath,amsfonts,amssymb}
\usepackage{subfig}
\usepackage{listings}
\usepackage{paralist}
\newcommand{\nospellcheck}[1]{#1}

\usepackage[x11names]{xcolor}
\usepackage{todonotes}

\newcommand{\push}{\texttt{push()}}
\newcommand{\pull}{\texttt{pull()}}
\newcommand{\pushfast}{\texttt{push\-Fast()}}
\newcommand{\pullfast}{\texttt{pull\-Fast()}}
\newcommand{\waitfor}{\texttt{wait\-For()}}

\newcommand{\load}{\ensuremath{\lambda}}

\newcommand{\ie}{\textit{i.e.}}
\newcommand{\etal}{\textit{et al.}}

\newcommand{\tr}[1]{%
\ensuremath{\overline{#1}}%
}
\newcommand{\dom}[1]{%
\ensuremath{\mathbb{#1}}%
}

\graphicspath{{fig/}}

\begin{document}

\maketitle
\begin{abstract}
Current runtime verification tools seldom make use of multi-threading to speed up the evaluation of a property on a large event trace. In this paper, we present an extension to the BeepBeep~3 event stream engine that allows the use of multiple threads during the evaluation of a query. 
Various parallelization strategies are presented and described on simple examples. The implementation of these strategies is then evaluated empirically on a sample of problems. 
Compared to the previous, single-threaded version of the BeepBeep engine, the allocation of just a few threads to specific portions of a query provides dramatic improvement in terms of running time.

\end{abstract}



%

\newsavebox{\cpuloadtable}
\begin{lrbox}{\cpuloadtable}
\begin{tabular}{|c|c|c|}
\hline
\textbf{Query} & \textbf{Threading Strategy} & \textbf{Load}\\
\hline\hline
{\href{T1.6.0}{Auction bidding}} & {\href{T1.7.1}{Multi-threaded}} & {\href{T1.7.2}{0.3446289}}\\
\cline{2-3}
 & {\href{T1.6.1}{None}} & {\href{T1.6.2}{0.60107625}}\\
\hline
{\href{T1.8.0}{Candidate selection}} & {\href{T1.9.1}{Multi-threaded}} & {\href{T1.9.2}{0.90646726}}\\
\cline{2-3}
 & {\href{T1.8.1}{None}} & {\href{T1.8.2}{0.3154293}}\\
\hline
{\href{T1.2.0}{Endless bashing}} & {\href{T1.3.1}{Multi-threaded}} & {\href{T1.3.2}{0.59272134}}\\
\cline{2-3}
 & {\href{T1.2.1}{None}} & {\href{T1.2.2}{0.3742459}}\\
\hline
{\href{T1.4.0}{Spontaneous Pingu creation}} & {\href{T1.5.1}{Multi-threaded}} & {\href{T1.5.2}{0.46694702}}\\
\cline{2-3}
 & {\href{T1.4.1}{None}} & {\href{T1.4.2}{0.30757308}}\\
\hline
{\href{T1.1.0}{Turn around}} & {\href{T1.1.1}{Multi-threaded}} & {\href{T1.1.2}{0.7721384}}\\
\cline{2-3}
 & {\href{T1.0.1}{None}} & {\href{T1.0.2}{0.35259247}}\\

\hline
\end{tabular}
\end{lrbox}

\newsavebox{\throughputtable}
\begin{lrbox}{\throughputtable}
\begin{tabular}{|c|c|c|}
\hline
\textbf{Query} & \textbf{Threading Strategy} & \textbf{Throughput}\\
\hline\hline
{\href{T2.7.0}{Auction bidding}} & {\href{T2.7.1}{Multi-threaded}} & {\href{T2.7.2}{30487.805}}\\
\cline{2-3}
 & {\href{T2.6.1}{None}} & {\href{T2.6.2}{28449.502}}\\
\hline
{\href{T2.9.0}{Candidate selection}} & {\href{T2.9.1}{Multi-threaded}} & {\href{T2.9.2}{127.93132}}\\
\cline{2-3}
 & {\href{T2.8.1}{None}} & {\href{T2.8.2}{28.794949}}\\
\hline
{\href{T2.2.0}{Endless bashing}} & {\href{T2.3.1}{Multi-threaded}} & {\href{T2.3.2}{742.4517}}\\
\cline{2-3}
 & {\href{T2.2.1}{None}} & {\href{T2.2.2}{511.77072}}\\
\hline
{\href{T2.4.0}{Spontaneous Pingu creation}} & {\href{T2.5.1}{Multi-threaded}} & {\href{T2.5.2}{782.4046}}\\
\cline{2-3}
 & {\href{T2.4.1}{None}} & {\href{T2.4.2}{742.0844}}\\
\hline
{\href{T2.1.0}{Turn around}} & {\href{T2.1.1}{Multi-threaded}} & {\href{T2.1.2}{30.937191}}\\
\cline{2-3}
 & {\href{T2.0.1}{None}} & {\href{T2.0.2}{16.566013}}\\

\hline
\end{tabular}
\end{lrbox}

\newsavebox{\throughputAndLoad}
\begin{lrbox}{\throughputAndLoad}
\begin{tabular}{|c|c|c|c|}
\hline
\textbf{Query} & \textbf{Threading Strategy} & \textbf{Throughput} & \textbf{Load}\\
\hline\hline
{\href{T3.6.0}{Auction bidding}} & {\href{T3.7.1}{Multi-threaded}} & {\href{T3.7.2}{30,487.8}} & {\href{T3.7.3}{0.345}}\\
\cline{2-4}
 & {\href{T3.6.1}{None}} & {\href{T3.6.2}{28,449.5}} & {\href{T3.6.3}{0.601}}\\
\hline
{\href{T3.9.0}{Candidate selection}} & {\href{T3.9.1}{Multi-threaded}} & {\href{T3.9.2}{127.9}} & {\href{T3.9.3}{0.906}}\\
\cline{2-4}
 & {\href{T3.8.1}{None}} & {\href{T3.8.2}{28.8}} & {\href{T3.8.3}{0.315}}\\
\hline
{\href{T3.2.0}{Endless bashing}} & {\href{T3.3.1}{Multi-threaded}} & {\href{T3.3.2}{742.5}} & {\href{T3.3.3}{0.593}}\\
\cline{2-4}
 & {\href{T3.2.1}{None}} & {\href{T3.2.2}{511.8}} & {\href{T3.2.3}{0.374}}\\
\hline
{\href{T3.4.0}{Spontaneous Pingu creation}} & {\href{T3.5.1}{Multi-threaded}} & {\href{T3.5.2}{782.4}} & {\href{T3.5.3}{0.467}}\\
\cline{2-4}
 & {\href{T3.4.1}{None}} & {\href{T3.4.2}{742.1}} & {\href{T3.4.3}{0.308}}\\
\hline
{\href{T3.1.0}{Turn around}} & {\href{T3.1.1}{Multi-threaded}} & {\href{T3.1.2}{30.9}} & {\href{T3.1.3}{0.772}}\\
\cline{2-4}
 & {\href{T3.0.1}{None}} & {\href{T3.0.2}{16.6}} & {\href{T3.0.3}{0.353}}\\

\hline
\end{tabular}
\end{lrbox}


\newcommand{\avgLoadincrease}{\href{M1.0}{1.7}}

\newcommand{\maxLoadincrease}{\href{M1.1}{2.9}}

\newcommand{\machinestring}{\href{M3.0}{AMD Athlon II X4 640 1.8 GHz running Ubuntu 16.04}}

\newcommand{\jvmram}{\href{M3.1}{898}}

\newcommand{\numexperiments}{\href{M3.2}{10}}

\newcommand{\numdatapoints}{\href{M3.3}{20}}

\newcommand{\avgThroughputincrease}{\href{M2.0}{2.0}}

\newcommand{\maxThroughputincrease}{\href{M2.1}{4.4}}

\section{Introduction}\label{sec:intro} 

Since its inception, the field of Runtime Verification (RV) has undergone a significant growth, both in the expressiveness of its specification languages, and in the range of problems it has addressed. Beyond finite-state machines and propositional temporal logics such as LTL, many runtime monitoring systems frequently support specification languages extending these formalisms in various ways. Moreover, it is not uncommon for RV use cases to involve systems generating tens of thousands of events per second. This puts pressure on the capacity of existing systems to efficiently process this flow of data, leading to the development of numerous optimization techniques.

Among these techniques, the leveraging of parallelism in existing computer systems has seldom been studied. Indeed, the prospect of parallel processing of temporal constraints in general, and LTL formul\ae{} in particular, is held back precisely because of the sequential nature of the properties to verify: since the validity of an event may depend on past and future events, the handling of parts of the trace in parallel and independent processes seems to be disqualified at the onset. Case in point, a review of available solutions in Section \ref{sec:related} of this paper observes that most existing trace validation tools are based on algorithms that do not take advantage of parallelism. While a few solutions do make use of threads, GPUs, distributed infrastructures or dedicated hardware, in virtually all cases parallelism is not employed for the evaluation of the property itself.

Current popular trace analysis software, such as the entrants to the latest Competition on Runtime Verification (CRV) \cite{DBLP:conf/rv/RegerHF16}, are globally single-thread tools. This is certainly true of the authors' BeepBeep 3, described in Section \ref{sec:bb3}, whose version entered at CRV 2016 did not use any multi-threading. Analysis of the available source code for MarQ \cite{RegerCR15}, the single other system entered in the competition's offline track, did not reveal the use of any multi-threading either. This is supported by visual inspection of the CPU load graph during the execution of the tool, which shows a high load on a single of the available cores at any point in time. 

\begin{figure}
\centering
\includegraphics[width=3.5in]{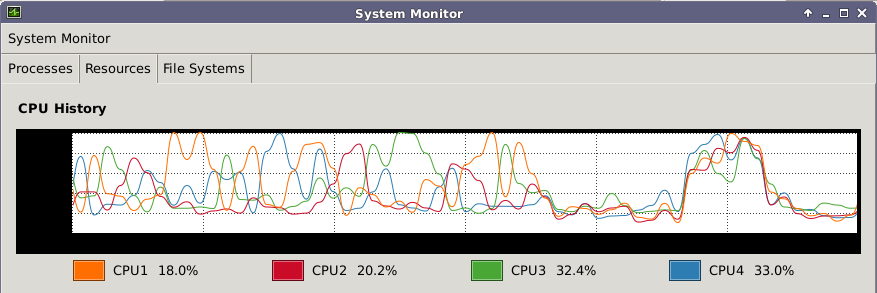}
\caption{An actual CPU load graph from the BeepBeep~3 system evaluating a query on a large event trace. The left part of the graph shows the query in single-threaded mode; the right part shows the same query running with multi-threading enabled.}
\label{fig:cpu-load-ordi}
\end{figure}

Figure \ref{fig:cpu-load-ordi} gives an example of this situation. The execution of the single-thread system is shown in the left part of the graph. One can see that, while the high CPU usage alternates between the four available cores of the host machine (a result of the operating system's load balancing), only one at a time exhibits a high load. This results in much available CPU power that is not used to speed up the system. On the contrary, a system harnessing all the available computing resources would result in a load graph similar to the right part of Figure \ref{fig:cpu-load-ordi}: this time, all four cores are used close to their full capacity. This figure also presents visual confirmation that the property, in this latter case, takes much less time to evaluate than in the single-thread scenario.

In this paper, we present a set of patterns that allow the use of multiple threads in the evaluation of queries on traces of events. Section \ref{sec:strategies} shows how these techniques leverage the architecture of the BeepBeep 3 event stream processor, whose computation of queries is implemented as the composition of multiple simple transducers called \emph{processors}. The techniques we propose are based on the idea of  ``wrapping'' a processor or a group of processors into one of a few special, thread-aware containers. This has for effect that any part of an event stream query can be parallelized, without requiring the wrapped processors to be aware of the availability of multiple threads. Thanks to this architecture, any existing computation can be made to leverage parallelism with very minimal modifications to the original query.

Section \ref{sec:implementation} then shows experimentally the potential impact of parallelism. Based on a sample of properties taken from the latest Competition on Runtime Verification (CRV), it reveals that five out of six queries show improved throughput with the use of parallelism, sometimes running as much as \maxThroughputincrease{} times faster.


\section{Parallelism in Runtime Verification}\label{sec:related} 

A distinction must be made between the runtime verification \emph{of} parallel and concurrent systems \cite{DBLP:journals/sttt/QadeerT12,DBLP:conf/sigsoft/SenRA03,DBLP:journals/tocs/SavageBNSA97,DBLP:conf/issta/LuoR13,DBLP:conf/spin/Harrow00,DBLP:conf/ifm/NazarpourFBBC16} and the use of parallelism \emph{for} runtime verification. This paper is concerned with the latter question.
Related literature can be split in two families, depending on what is being parallelized. We shall call these two families ``second-hand'' and ``first-hand'' parallelism. 

\subsection{Second-Hand Parallelism}

A first set of works provide improvements to elements that are peripheral to the evaluation of the property, such as the communication between the monitor and the program, or the use of a separate chip to run the monitor. We call this form of parallelism ``second-hand''. The prospect of using physical properties of hardware to boost the performance of runtime verification has already been studied in the recent past. For example, Pellizzoni \etal\@ \cite{pellizzoni2008hardware} utilized dedicated commercial-off-the-shelf (COTS) hardware \cite{emerson1990temporal} to facilitate the runtime monitoring of critical embedded systems whose properties were expressed in Past-time Temporal Linear Logic (PTLTL).

As the number of cores (GPU or multi-core CPUs) in the commodity hardware keeps increasing, the research of exploiting the available processors or cores to parallelize the tasks and the computing  brings a challenge and also an opportunity to improve the architecture of runtime verification. For example, Ha \etal\@ \cite{ha2009concurrent} introduced a buffering design called Cache-friendly Asymmetric Buffering (CAB) to improve the communications between application and runtime monitor by exploiting the shared cache of the multi-core architecture; Berkovich \etal\@ \cite{DBLP:journals/fmsd/BerkovichBF15} proposed a GPU-based solution that effectively utilizes the available cores of the GPU, so that the monitor designed and implemented with their method can run in parallel with the target program and evaluate LTL properties.

\subsection{First-Hand Parallelism}

While the previous works all claim improvements in the efficiency of the monitoring process, they do not address directly the issue of performing the evaluation of the property itself using parallelism. Therefore, a second family of related works address the issue of what we shall call ``first-hand'' parallelism. In various ways, these works attempt to split the computation steps of evaluating a query on a trace into blocks that can be executed independently over multiple processes.

\paragraph{In Runtime Verification}

When trace properties are expressed as Linear Temporal Logic, 
Kuhtz and Finkbeiner showed that the path checking problem (\ie\@ verifying that a given trace fulfills the property) belongs to the complexity class \textsf{AC$^1$(logDCFL)} \cite{DBLP:journals/corr/abs-1210-0574}. This result entails that the process can be efficiently split by evaluating entire blocks of events in parallel. Rather than sequentially traversing the trace, their work considers the circuit that results from ``unrolling'' the formula over the trace. However, while the evaluation of this unrolling can be done in parallel, a specific type of Boolean circuit requires to be built in advance, which depends on the length of the trace to evaluate. Moreover, the argument consists of a formal proof; deriving a concrete, usable implementation of this technique remains an open question.

In an offline context, the leveraging of a cluster infrastructure was first suggested by Hallé \etal, who introduced an algorithm for the automated verification of Linear Temporal Logic formul\ae{} on event traces, using an increasingly popular cloud computing framework called MapReduce \cite{jocasa-2015}. The algorithm can process multiple, arbitrary fragments of the trace in parallel, and compute its final result through a cycle of runs of MapReduce instances.
A different architecture using MapReduce was proposed by Basin \etal\@ \cite{DBLP:journals/fmsd/BasinCEHKM16}. These techniques only apply in a batch (\ie\@ offline) context. Moreover, the setup of a MapReduce infrastructure is a heavy process that only pays off for massive amounts of data to be evaluated against the same property. At least in the case of the first approach, it was discovered experimentally that the anticipated speed-up caused by parallelism was offset by the increased amount of communication and the sheer volume of tuples that needed to be manipulated.

A second downside of these approaches is that both are tailored to one specific query language ---in both cases an extension of Linear Temporal Logic. It remains unclear how computing other types of results over an event trace could be done in the same framework.

\paragraph{In Event Stream Processing} Much more work on parallelism and distribution was undertaken from the database point of view. The evaluation of SQL queries using multiple cores has spawned a large line of works \cite{DBLP:conf/vecpar/PaesLVM08,DBLP:conf/vldb/KrikellasVC10,DBLP:conf/ISCApdcs/LiKNS06,DBLP:conf/sigmod/GangulyHK92,DBLP:journals/tkde/ApersBFGKW92,DBLP:reference/db/Graefe09d}. However, relational database queries differ significantly from runtime verification properties, where the order of events is an essential matter. This rules out the possibility of a simple sharding of the source data into independents partitions that can be processed independently, as sequential relationships between events across multiple shards can go missing.

Much closer to RV's concerns is the field of event stream processing (ESP). An ESP system generally consists of one or more unbounded sequences of events (the streams), which are sent to processing units connected according to a directed acyclic graph (DAG). Such a graph, also called a \emph{query}, can run continuously on incoming streams or by reading pre-recorded data. The similarities between runtime verification and ESP have been highlighted in a recent tutorial \cite{DBLP:conf/rv/Halle16}.

Some recent ESP systems, like Siddhi \cite{Suhothayan:2011:SSL:2110486.2110493}, appear to be single-machine, single-thread query engines. However, many others naturally support processing units to be distributed on multiple machines that exchange data through a network communication. By manually assigning fragments of the query to different machines, distribution of computation can effectively be achieved. Such systems include Aurora, Borealis \cite{DBLP:conf/cidr/AbadiABCCHLMRRTXZ05}, Cayuga \cite{DBLP:conf/sigmod/BrennaDGHOPRTW07}, Apache Storm \cite{storm}, Apache S4 \cite{DBLP:conf/icdm/NeumeyerRNK10} and Apache Samza \cite{samza}. This form of parallelism is called ``inter-host'', as the computation is performed on physically distinct devices.

Some systems also support ``intra-host'' parallelism, this time by allotting fragments of a task to multiple cores (or threads) within a single host. One notable example is Esper \cite{esper}, which has documented multi-threading facilities. Of special interest is a system called PIPES, which provides a three-level multi-threaded architecture \cite{Cammert03pipes:a}. On the first level, operators are directly connected, forming virtual operators; on the second level, virtual operators are combined inserting buffers in-between. Separate threads control the so formed subgraphs. Finally, on the top level, a scheduler enables concurrent execution of these subgraphs.


\section{The BeepBeep 3 Event Stream Query Engine}\label{sec:bb3} 

BeepBeep~3 is an event stream processing engine. It can can be used either as a Java library embedded in another application's source code, or as a stand-alone query interpreter running from the command-line. Releases of BeepBeep~3 are publicly available for download under an open source license.\footnote{\url{https://liflab.github.io/beepbeep-3}} BeepBeep can be used either as a Java library embedded in another application's source code, or as a stand-alone query interpreter running from the command-line. In this section, we briefly describe the basic principles underlying the architecture of BeepBeep. The reader is referred to a recent tutorial for more details, such as the complete formal semantics of the language and some examples \cite{DBLP:conf/rv/Halle16}.

\subsection{Processors}

Let $\dom{T}$ be an arbitrary set of elements. An \emph{event trace of type $\dom{T}$} is a sequence $\tr{e} = e_0 e_1 \dots$ where $e_i \in \dom{T}$ for all $i$. Event types can be as simple as single characters or numbers, or as complex as matrices, XML documents, plots, logical predicates, polynomials or any other user-defined data structure.

A \emph{function} is an object that takes zero or more events as its input, and produces zero or more events as its output. In BeepBeep, functions are first-class objects; they all descend from an abstract ancestor named \texttt{Function}, which declares a method called \texttt{evaluate()} so that outputs can be produced for a given array of inputs. 
A \emph{processor} is an object that takes zero or more event \emph{traces}, and produces zero or more event \emph{traces} as its output. A function is stateless, and operates on individual events: given an input, it immediately produces an output, and the same output is always returned for the same inputs. On the contrary, 
a processor is a stateful device: for a given input, its output may depend on events received in the past. Any processors with compatible types can be freely composed.

A processor produces its output in a \emph{streaming} fashion: it does not wait to read its entire input trace before starting to produce output events. However, a processor can require more than one input event to create an output event, and hence may not always output something when given an input. Processors can then be composed (or ``piped'') together, by letting the output of one processor be the input of another. When a processor has an input arity of 2 or more, the processing of its input is done \emph{synchronously}. This means that a computation step will be performed if and only if an event can be consumed from each input trace. 
This entails that processors must implicitly manage \emph{buffers} to store input events until a result can be computed. This buffering is implicit: it is absent from both the formal definition of processors and any graphical representation of their piping. Nevertheless, the concrete implementation of a processor must take care of these buffers in order to produce the correct output. 
In BeepBeep, this is done with the abstract class \texttt{SingleProcessor}; descendents of this class simply need to implement a method named \texttt{compute()}, which is called only when an event is ready to be consumed at each input.

\subsection{Built-In Processors and Palettes}

BeepBeep is organized along a modular architecture. The main part of BeepBeep is called the \emph{engine}, which provides the basic classes for creating processors and functions, and contains a handful of general-purpose processors for manipulating traces. The rest of BeepBeep's functionalities is dispersed across a number of optional \emph{palettes}, which will be discussed later.

A first way to create a processor is by lifting any $m:n$ function $f$ into a $m:n$ processor. This is done by applying $f$ successively to each input event, producing the output events. A few processors can be used to alter the sequence of events received. The \texttt{CountDecimate} processor returns every \nospellcheck{$n$-th} input event and discards the others. 
Another operation that can be applied to a trace is trimming its output. Given a trace, the \texttt{Trim} processor returns the trace starting at its \nospellcheck{$n$-th} input event.

Events can also be discarded from a trace based on a condition. The \texttt{Filter} processor \textsc{f} is a $n:n-1$; the events are let through on its $n-1$ outputs, if the corresponding event of input trace $n$ is the Boolean value true ($\top$); otherwise, no output is produced.

BeepBeep also allows users to define their own processors directly as Java objects, using no more than a few lines of boilerplate code. The simplest way to do so is to extend the \texttt{SingleProcessor} class, which takes care of most of the ``plumbing'' related to event management: connecting inputs and outputs, looking after event queues, etc. All that is left to do is to define its input and output arity, and to write the actual computation that should occur, i.e.\ what output event(s) to produce (if any), given an input event. A detailed description of this extension mechanism has recently been published \cite{bigdata2016}.


\section{Multi-Threading Patterns}\label{sec:strategies} 

In this section, we now present a set of \emph{patterns} that can be applied to processors to make use of multiple threads. Special care has been given to make these multi-threading capabilities transparent. For the developer, this means that new processors can be created in a completely sequential manner; multi-threading can be applied at a later stage by simply wrapping them into special constructs that make use of threads. For the user, this means that queries are constructed in the same way, whether multi-threading is used or not: multi-threaded elements have the same interface as regular ones, and both types of elements can be freely mixed in a single query.

\subsection{Thread Management Model}

A first feature is the global thread management model. In BeepBeep, the instantiation of all threads is controlled by one or more instances of the \verb+ThreadManager+ class. Each thread manager is responsible for the allocation of a pool of threads of bounded size. A processor wishing to obtain a new thread instance asks permission from its designated thread manager. If the maximum number of live threads reserved to this manager is not exceeded, the manager provides a new thread instance to the processor. The processor can then associate any task to this thread and start it.

So far, this scheme is identical to traditional thread pooling found in various systems, or, in an equivalent manner, to the \verb+Executor+ pattern found in recent versions of Java. The main difference resides in the actions taken when the manager refuses to provide a new thread to the caller. Typically, the call for a new thread would become blocking until a thread finally becomes available. In BeepBeep, on the contrary, the call simply returns \verb+null+; this indicates that whatever actions the processor wished to execute in parallel in a new thread should instead be executed sequentially within the \emph{current} thread.

In the extreme case, when given a thread manager that systematically refuses all thread creation, the processor's operation is completely sequential; no separate thread instance is ever created.\footnote{Note that this is different from a thread manager that would dispense only one thread at a time. In such a case, \emph{two} threads are active: the calling thread and the one executing the task.} 
Moreover, various parts of a query can be assigned to different thread manager instances with their own thread pool, giving flexible control over how much parallelism is allowed, and to what fragments of the computation. Hence the amount of threading can be easily modulated, or even completely turned off quite literally at the flick of a switch, by changing the value of a single parameter. Since thread instances are requested dynamically and are generally short-lived, this means that changes in the threading strategy can also be made during the processing of an event stream.

The upper cap on the number of threads imposed by each thread manager only applies to \emph{running} threads. Hence if a set of 10 tasks $T_1, \dots, T_{10}$ it to be started in parallel using a thread manager with a limit of 3, the first three tasks will each be started in a new thread. However, if any of these threads finishes before requesting a thread for $T_4$, then $T_4$ will also be granted a new thread; the same applies for the remaining tasks. Finished threads are kept by the thread manager until their \texttt{dispose()} method is called, indicating they can be safely destroyed.

Note that while this technique can be compared to thread scheduling, it is an arguably simpler form of scheduling. A thread is either created or not, depending solely on whether the maximum number of live threads assigned to the manager has been reached at the moment of the request. The balancing of threads across various parts of a query graph is achieved by assigning these different parts to a different manager.

\subsection{Non-Blocking Push/Pull}

In a classical, single-thread computation, a call to any of a processor's Pullable or Pushable methods is \emph{blocking}. For example, calling  \pull{} on one of a processor's Pushables involves a call to the processor's input pullables \pull{} in order to retrieve input events, followed by the computation of the processor's output event. The original call returns only when this chain of operations is finished. The same is true of all other operations.

The Pushable and Pullable interfaces also provide a non-blocking version of the push and pull operations, respectively called \pushfast{} and \pullfast{}. These methods perform the same task as their blocking counterparts, but \emph{may} return control to the caller before the operation is finished. In order to ``catch up'' with the end of the operation, the caller must, at a later time, call method \waitfor{} on the same instance. This time, this method blocks until the push or pull operation that was started is indeed finished.

Following the spirit of transparency explained earlier, \pushfast{} and \pullfast{} are not required to be non-blocking. As a matter of fact, the default behaviour of these two methods is to act as a proxy to their blocking equivalents; similarly, the default behaviour of \waitfor{} is to return immediately. Thus, for a processor that does not wish to implement non-blocking operations, calling e.g.\ \pushfast{} followed by \waitfor{} falls back to standard, blocking processing. Only when a processor wishes to implement special non-blocking operations does it need to override these defaults.

Rather than implementing non-blocking operations separately for each type of processor, an easier way consists of enclosing an existing processor within an instance of the \texttt{Non\-Blocking\-Processor} class. When \push{} or \pull{} is called on this processor, a new thread is asked to its designated thread manager. The actual call to \push{} (resp.\ \pull{}) on the underlying processor is started in that thread and the control is immediately returned to the caller. Using a simple Boolean semaphore, a call to method \waitfor{} of the \texttt{Non\-Blocking\-Processor} sleep-loops until that thread stops, indicating that the operation is finished. We remind the reader that the thread manager may not provide a new thread; in such a case, the call to the underlying processor is made within the current thread, and the processor falls back to a blocking mode.

Non-blocking push/pull does not provide increased performance in itself. As a matter of fact, calling \pushfast{} immediately followed by \waitfor{} may end up being slower than simply calling \push{}, due to the overhead of creating a thread and watching its termination. However, it can prove useful in situations where one calls \push{} or \pull{} on a processor, performs some other task $T$, and retrieves the result of that \push{} (resp.\ \pull{}) at a later time. If the call is done in a non-blocking manner, the computation of that operation can be done in parallel with the execution of $T$ in the current thread.

It turns out that a couple of commonly used processors in BeepBeep's palettes operate in this fashion, and can hence benefit from the presence of non blocking push/pull methods. We describe a few.

\paragraph{Sliding Window}

Given a processor $\varphi$ and a window width $n$, the sliding window processor returns the output of a copy of $\varphi$ on events $e_0 e_1 \dots e_{n-1}$, followed by the output of a fresh copy of $\varphi$ on events $e_1 e_2 \dots e_{n}$, and so on. One possible implementation of this mechanism is to keep in memory up to $n-1$ copies of $\varphi$, such that copy $\varphi_i$ has been fed the last $i$ events received. Upon processing a new input event, the window pushes this event to each of the $\varphi_i$, and retrieves the output of $\varphi_{n-1}$. This processor copy is then destroyed, the index of the remaining copies is incremented by 1, and a new copy $\varphi_0$ is created.

\begin{figure}
\centering
\subfloat[Blocking]{\includegraphics[scale=0.8]{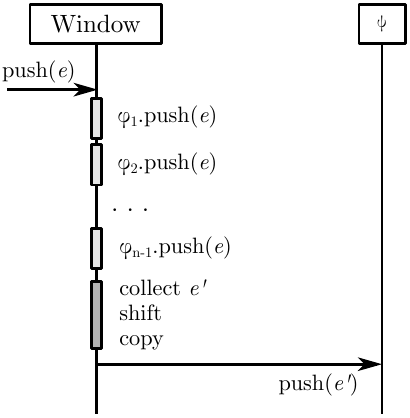}\label{subfig:window-sequential}}
\hspace{1cm}
\subfloat[Non-blocking]{\includegraphics[scale=0.8]{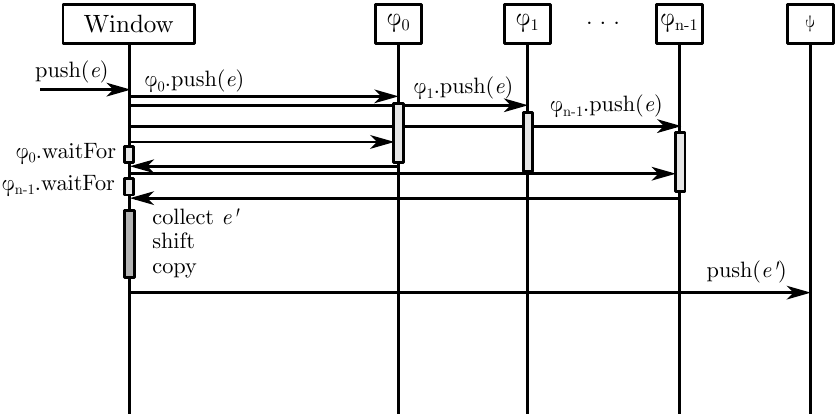}\label{subfig:window-parallel}}
\caption{Sequence diagram for the Window processor: (a) using blocking calls to $\varphi$; (b) using non-blocking calls running in separate threads.}
\label{fig:window}
\end{figure}

Figure \ref{subfig:window-sequential} shows a sequence diagram of these operations when performed in a blocking way. Figure \ref{subfig:window-parallel} shows the same operations, this time using non-blocking calls. The window processor first calls \push{} on each copy in rapid-fire succession. Each copy of $\varphi$ can update its state in a separate thread, thus exhibiting parallel processing. The processor then waits for each of these calls to be finished, by calling \waitfor{} on each copy of $\varphi$. The remaining operations are then performed identically.

This figure provides graphical evidence that, under the assumption that each call to \push{} occurs truly in parallel, the total time spent on the window's \push{} call is shorter than its sequential version. If $T_\varphi$ is the time for each call to $\varphi$ and $T_C$ is the time taken for the remaining tasks, then the call to this method goes down from $nT_\varphi + T_C$ to $T_\varphi + T_C$. If fewer than $n$ threads are available, the value is situated somewhere between these two bounds.

\paragraph{Trace Slicing}

The \texttt{Slicer} is a 1:1 processor that separates an input trace into different ``slices''. It takes as input a processor $\varphi$ and a function $f : \mathbb{T} \rightarrow \mathbb{U}$, called the \emph{slicing function}. There exists potentially one instance of $\varphi$ for each value in the image of $f$. If $\mathbb{T}$ is the domain of the slicing function, and $\mathbb{V}$ is the output type of $\varphi$, the slicer is a processor whose input trace is of type $\mathbb{T}$ and whose output trace is of type $2^\mathbb{V}$.

When an event $e$ is to be consumed, the slicer evaluates $c = f(e)$. This value determines to what instance of $\varphi$ the event will be dispatched. If no instance of $\varphi$ is associated to $c$, a new copy of $\varphi$ is initialized. Event $e$ is then given to the appropriate  instance of $\varphi$. Finally, the last event output by every instance of $\varphi$ is collected into a set, and that set is the output event corresponding to input event $e$. The function $f$ may return a special value $\#$, indicating that no new slice must be created, but that the incoming event must be dispatched to \emph{all} slices. In this latter case, a task similar to the Window processor can be done: each slice is put in a separate thread, so that it can process the input event in parallel with the others.

\paragraph{Logical Connectives}

BeepBeep comes with a palette providing processors for evaluating all operators of Linear Temporal Logic (LTL), in addition to the first-order quantification defined in LTL-FO$^+$ (and present in previous versions of BeepBeep) \cite{DBLP:journals/tsc/HalleV12}. Boolean processors are called \texttt{Globally}, \texttt{Eventually}, \texttt{Until}, \texttt{Next}, \texttt{ForAll} and \texttt{Exists}, and carry their usual meaning. For example, if $a_0 a_1 a_2 \dots$ is an input trace, the processor \texttt{Globally} produces an output trace $b_0 b_1 b_2 \dots$ such that $b_i = \bot$ if and only there exists $j \geq i$ such that $b_j = \bot$. In other words, the $i$-th output event is the two-valued verdict of evaluating $\mbox{\bf G}\,\varphi$ on the input trace, starting at the $i$-th event.\footnote{Another set of processors, called the ``Trooleans'', mirror each of these operators but use a three-valued semantics. The reader is referred to the BeepBeep 3 tutorial for more details on this \cite{DBLP:conf/rv/Halle16}.}

Concretely, this is implemented by creating one copy of the $\varphi$ processor upon each new input event. This event is then pushed into all the current copies, and their resulting output event (if any) is collected. If any of them is $\bot$ (false), the processor returns $\bot$; if any of them is $\top$ (true), the corresponding copy of $\varphi$ is destroyed. This is another example where the processing of multiple copies of $\varphi$ can be done in a separate thread, in a way similar to the principles described earlier. The same can be done with first-order quantifiers; hence for the expression $\forall x \in \pi : \varphi(x)$, the evaluation of $\varphi$ for each value of $x$ can be done in parallel.

\subsection{Pre-emptive Pulling}

A second strategy consists of continuously pulling for new outputs on a processor $P$, and storing these events in a queue. When a downstream processor $P'$ calls $P's$ \pull{} method, the next event is simply popped from that queue, rather than being computed on-the-fly. If $P'$ is running in a different thread from the process that polls $P$, each can compute a new event at the same time.

\begin{figure}
\centering
\subfloat[Sequential]{\includegraphics[scale=0.8]{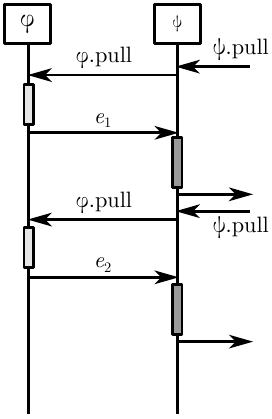}\label{subfig:pull-sequential}}
\hspace{1cm}
\subfloat[Pre-emptive pulling]{\includegraphics[scale=0.8]{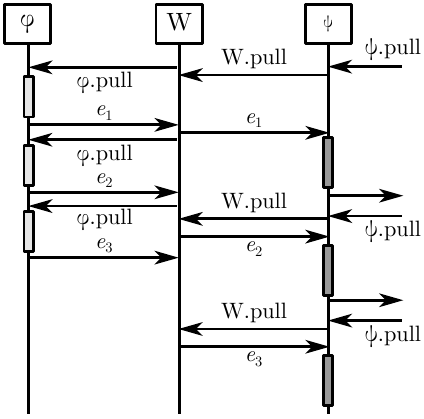}\label{subfig:pull-parallel}}
\caption{Sequence diagram for pre-emptive pulling: (a) no pre-emptive pulling; (b) W performs pre-emptive pulling on $\varphi$ in a separate thread.}
\label{fig:pull}
\end{figure}

Figure \ref{subfig:pull-sequential} shows the processing of an event when done in a sequential manner. A call to \pull{} on $\psi$ results in a pull on $\varphi$, which produces some output event $e$. This event is then processed by $\psi$, which produces some other output $e'$. If $T_\varphi$ and $T_{\psi}$ correspond to the computation times of $\varphi$ and $\psi$, respectively, then the total time to fetch each event from $\psi$ is their sum, $T_\varphi + T_{\psi}$.

On the contrary, Figure \ref{subfig:pull-parallel} shows the same process, with pre-emptive pulling on $\varphi$ occurring in a separate thread. One can see that in this case, $\varphi$ produces a new output event while $\psi$ is busy doing its computation on the previous one. The first output event still takes $T_\varphi + T_{\psi}$ to be produced, but later ones can be retrieved in $\max\{T_\varphi, T_{\psi}\}$.

In a manner similar to the \texttt{Non\-Blocking\-Processor}, pre-emptive pulling is enabled by enclosing a group of processors inside a \texttt{PullGroup}. This processor behaves like a \texttt{GroupProcessor}: a set of connected processors can be added to the group, and this group can then be manipulated, piped and copied as if it were a single ``black box''. The difference is that a call to the \texttt{start()} method of a \texttt{PullGroup} creates a new thread where the repeated polling of its outputs occurs. To avoid needlessly producing too many events that are not retrieved by downstream calls to \pull{}, the polling stops when the queue reaches a predefined size; polling resumes when some events of that queue are finally pulled. As with the other constructs presented in this paper, the \texttt{PullGroup} takes into account the possibility that no thread is available; in such a case, output events are computed only upon request, like the regular \texttt{GroupProcessor}.

In our analysis of computing time, we can see that the speed gain is maximized when $T_\varphi = T_\psi$; otherwise, either $\varphi$ produces events faster than $\psi$ can consume them ($T_\varphi < T_\psi$), or $\psi$ wastes time waiting for the output of $\varphi$ ($T_\varphi > T_\psi$). Therefore, an important part of using this strategy involves breaking a processor graph into connected regions of roughly equal computing load.

\subsection{Pipelining}

Pipelining is the process of reading $n$ input events $e_1 e_2 \dots e_n$, creating $n$ copies of a given processor, and launching each of them on one of the input events. A pipeline then waits until the processor assigned to $e_1$ produces output events; these events are made available at the output of the pipeline as they are collected. Once the $e_1$ processor has no more output events to produce, it is discarded, the collection resumes on the processor for $e_2$, and so on.

Note that, once the $e_1$ processor is discarded, there is now room for creating a new processor copy and assign it to the next input event, $e_{n+1}$. This rolling process goes on until no input event is available. In such a scheme, the order of the output events is preserved: in sequential processing, the batch of output of events produced by reading event $e_1$ comes before any output event resulting from processing $e_2$ is output.

\begin{figure}
\centering
\subfloat[Sequential]{\includegraphics[scale=0.8]{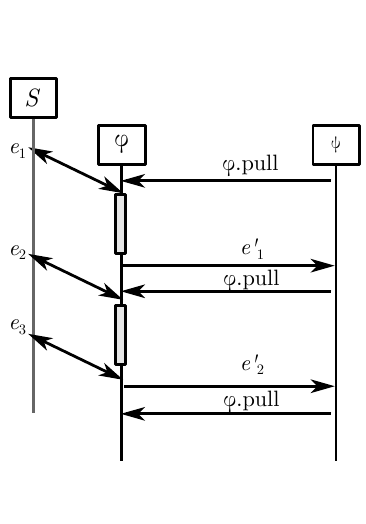}\label{subfig:pipeline-sequential}}
\hspace{1cm}
\subfloat[Pipelining]{\includegraphics[scale=0.8]{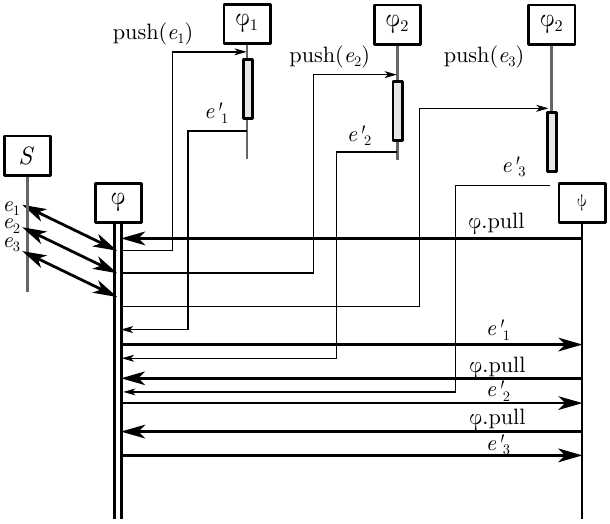}\label{subfig:pipeline-parallel}}
\caption{Sequence diagram for pipelining: (a) no pipelining: $\varphi$ requests events from $S$ on demand and computes its output event afterwards; (b) $\varphi$ pulls multiple events from $S$ and evaluates each of them on a copy of $\varphi$ in a separate thread.}
\label{fig:pipeline}
\end{figure}

Although pipelining borrows features from both pre-emptive pulling and non-blocking pull, it is actually distinct from these two techniques. As in non-blocking pull, it sends input events in parallel to multiple copies of a processor; however, rather than sending the same event to multiple, independent instances of $\varphi$, it sends events that should be processed in sequence by a single processor instance each to a distinct copy of $\varphi$ and collects their result. In the sequence $e_1 e_2 \dots$, this means that one copy of $\varphi$ processes the subtrace $e_1$, while another one processes (in parallel) the subtrace $e_2$, and so on.

Obviously, this ``trick'' does not guarantee the correct result on all processors, as some of them have an output that depends on the complete trace. 
As a simple example, one can consider the processor that computes the cumulative sum of all input events; it is clear that pipelining this processor will return an incorrect result, as each copy of the processor will receive (and hence add) a unique event of the input trace.
However, there do exist processors for which pipelining can be applied; this is the case of all \texttt{Function\-Processor}s, which by definition apply a stateless function to each of their input events. While this might seem limiting, it turns out that, in the sample of queries evaluated experimentally later in this paper, a large part of the computing load comes from the application of a few such functions, and that pipelining proves very effective in these situations.

We also remark that the process could be generalized in at least two ways. First, a stateful processor may become stateless after reading a trace prefix; for example, a processor that computes the sum of input events until it reaches 10, or a Moore machine that reaches a sink state. Therefore, an ``adaptive'' pipeline could query a processor for such a condition, and start pipelining once the processor becomes stateless. Second, some processors may be stateful, but only with respect to a bounded number $n$ of past events. If this bound is known, pipelining can be applied by sending to each copy of the processor the window of the last $n$ events; the pipeline presented here is the simple case where $n=1$.


\section{Implementation and Discussion}\label{sec:implementation} 

All the concepts described in the previous section have been implemented in the BeepBeep 3 event processing engine and are available in its latest downloadable version. In this section, we discuss some of the advantages of this architecture and present experimental results indicating the potential speedup it can bring.

\subsection{Experiments}

In order to showcase the usage (and potential advantages) of using multiple threads according to our proposed architecture, we setup a set of experiments where a modified version of BeepBeep is run on a set of example queries and input traces. The queries and input traces are taken from the Offline track of the 2016 Competition on Runtime Verification \cite{DBLP:conf/rv/RegerHF16}, in which the single-thread version of BeepBeep 3 had participated. For each property and each thread, BeepBeep is run in two configurations:
\begin{inparaenum}
\item The original query, without any thread-aware processors
\item The same query, modified with thread-aware processors inserted in the ``best'' way possible. These modifications were done with the help of intuition and manual trial and error, so they may not represent the absolute best way of using threads. The queries use a single thread manager, whose number of threads is set to be equal to the host machine's number of cores. No other modifications to the query have been made.
\end{inparaenum}

The experiment measures two elements. The first is \emph{throughput}, measured in Hz, and which corresponds to the average number of input events consumed per second. The main goal of multi-threading is to achieve faster computation; it is therefore natural to get a hint on the amount of speed-up one can get by involving more than one processing thread. The second is \emph{CPU load}, measured in percentage. For a system with $n$ cores, let $f_i(t)$ be a function giving the instantaneous load (between 0 and 1) of core $i$ at time $t$. If $T_S$ and $T_E$ are the start and end time of the execution of a query, then the load resulting from the execution of that query, noted \load{}, is defined as:
\[
\load \triangleq \frac{1}{n(T_E - T_S)} \sum_{i=1}^n \int_{T_S}^{T_E} f_i(t) dt
\]

Intuitively, the load represents how much of the available cores was used during the execution of the query. For example, a query resulting in a constant usage of 50\% on two cores of a 4-core machine and 0\% on the remaining two would have a load of 0.25. Note that load is a value between 0 and 1 that is not affected by the duration of the computation.
In the experiments, instantaneous load is approximated as a 1-second wide rectangle, whose height corresponds to the CPU usage as reported by the SIGAR API\footnote{\url{https://support.hyperic.com/display/SIGAR/Home}}. Note that this usage includes that of all applications running on the host machine. Extraneous activity was minimized by having the machine boot into runlevel 3 (command-line without X server) with a limited number of running services and daemons. 
%
%


As is now customary in LIF research projects, all experiments and data are available for download as a self-contained application that can be launched and controlled from a web interface. This bundle has been created using the LabPal testing library\footnote{\url{https://liflab.github.io/labpal}. The actual lab is available at \url{https://datahub.io/dataset/beepbeep-mt-lab}.}, and allows anyone to easily re-run the same queries on the same input data.

The results of these experiments are summarized in Table \ref{tab:experiments}. One of the queries (SQL Injection) is implemented in BeepBeep as a single finite-state machine that simply updates the contents of a set upon each incoming event. None of the patterns introduced in this paper can be applied to this processor, and so no experiment was conducted on it. (We remind the reader that our proposed solution does not claim to be universal.)

\begin{table}
\centering
\scalebox{0.85}{\usebox{\throughputAndLoad}}
\caption{Throughput and load for each query, for the single-thread and multi-thread versions of BeepBeep.}
\label{tab:experiments}
\vskip -0.5cm
\end{table}

As one can see, the use of multi-threading has increased the throughput and load of every remaining problem instance. CPU load has increased by a factor of \avgLoadincrease{}, and throughput by a factor of \avgThroughputincrease{}. In some cases, such as the Auction property, the impact of parallelism is dramatic. This is due to the fact that this property evaluates a first-order logic formula with multiple nested quantifiers, such as $\forall a \in A : \forall b \in B : \exists c \in C : \varphi$. When the cardinality of sets $B$ and $C$ is large, evaluating $\forall b \in B : \exists c \in C : \varphi$ for each value $a \in A$ can efficiently be done in parallel.

\subsection{Discussion}

We shall now discuss the pros and cons of the multi-threading architecture proposed in this paper.

\paragraph{Simplicity} A first advantage of the proposed architecture is its high simplicity. Using either of the multi-threading strategies described earlier can generally be done using a handful of instructions and very shallow modifications to an existing query.

For example, the piece of code in Figure \ref{fig:code-example} shows how to enable non-blocking push/pull within a window processor. The regular, mono-threaded version would include only lines 2 and 4: one creates a cumulative sum processor, which is then given to the Window processor with a specific window width (10 in this case). The multi-threaded version adds lines 1 and 3. The first instruction creates a new thread manager and gives it an upper limit of four threads (one could also reuse an existing manager instance instead of creating a new one). The third instruction wraps a non-blocking processor around the original sum processor; it is this processor that is given to the window instead of the original. The end result is that the calls to \texttt{sum}'s \push{} and \pull{} will be split across four threads, and that \texttt{w} will operate like in Figure \ref{subfig:window-parallel}.

\begin{figure}
\begin{lstlisting}
1 ThreadManager manager = new ThreadManager(4);
2 FunctionProcessor sum = new FunctionProcessor(new CumulativeFunction(Addition.instance));
3 NonBlockingProcessor nbp = new NonBlockingProcessor(sum, manager);
4 Window w = new Window(nbp, 10);
\end{lstlisting}
\caption{Enabling non-blocking calls inside a window processor for parallel processing.}
\label{fig:code-example}
\end{figure}

Note how adding multi-threading to this example involves very minimal modifications to the original query; it merely amounts to wrapping a processor instance around another one. Apart from the way it handles push/pull internally, that processor behaves exactly like the original. Thanks to the blocking fallback semantics of \pullfast{} and \pushfast{}, the window processor that uses it does not even need to be aware that it is handed a multi-threaded processor.

\paragraph{Separation of Concerns} This observation brings us to the second important benefit of this architecture: multi-threaded code remains very localized. For example, the \texttt{Window} processor does not contain any special code to handle multi-threading; the same can be said of all other processor objects. Hence a user is not required to take multi-threading into account when writing a new, custom processor. Parallelism occurs only by virtue of enclosing some processor instances inside thread-aware wrappers.

A telling indicator of this separation of concerns can be obtained from analyzing BeepBeep's code: outside the \texttt{concurrency} package that provides the objects introduced in this paper (the thread manager and a handful of processor wrappers, which amount to roughly 2,000 lines of code), no reference to threads is ever made across the rest of BeepBeep's code, including all its palettes (18,000 lines). As a matter of fact, BeepBeep can even be compiled with this package deleted without affecting any of its functionalities.

\paragraph{Manual Definition of Parallel Regions}

In counterpart, the current implementation of multi-threading in BeepBeep requires the user to explicitly define the regions of a query graph that are to be executed using multiple threads. This requires some knowledge of the underlying computations that are being executed in various parts of the query, and some intuition as to which of them could best benefit from the availability of more than one thread. Doing so improperly can actually turn out to be detrimental to the overall throughput of the system.

Therefore, the architecture proposed in this paper should be taken as a first step. It breaks new ground by providing a simple way to add multi-threading to the evaluation of an arbitrary query; however, in the medium term, higher-level techniques for selecting the best regions of a query suitable for parallelism should be developed. It shall be noted that this issue is not specific to BeepBeep, and that the automated fine tuning of parallelism in query evaluation is a long-standing research problem \cite{DBLP:conf/damon/MuhlbauerRSK014,DBLP:conf/icde/BelknapDDY09,DBLP:journals/tkde/Viglas14,DBLP:journals/tpds/GedikSHW14}. As an anecdotal proof, Oracle devotes a whole chapter of its documentation on intricate tuning tips for the evaluation of SQL queries \footnote{\url{http://docs.oracle.com/cd/A84870_01/doc/server.816/a76994/tuningpe.htm}}. Readers should not expect that BeepBeep, or any other monitor system, could transparently (one might say magically) setup threads in the best way for all possible queries.



\section{Conclusion} 

In this paper, we have introduced a few simple patterns for the introduction of parallelism in the evaluation of queries over streams of events. These patterns can be applied in systems whose computation is expressed as the composition of small units of processing arranged into a graph-like structure, as is the case in the BeepBeep 3 event stream engine. By surrounding appropriate regions of such a graph with special-purpose thread managers, parts of a query can be evaluated using multiple cores of a CPU, therefore harnessing more of a machine's computing power.

Thanks to the simplicity of these patterns and to BeepBeep's modular design, the introduction of parallelism to an existing query requires very limited changes, which amount to the insertion of two or three lines of code at most. 
To the best of our knowledge, this makes BeepBeep the first runtime verification tool with such parallelization capabilities. 
However, as we have seen, not all queries are amenable to parallelization, at least not using the patterns introduced in this paper. Therefore, the introduction of multi-threading should also be complemented with advances on other fronts; for example, recent works have shown how a clever use of internal data structures can also provide important speed gains \cite{DBLP:conf/tacas/DeckerHS0T16}.

Nevertheless, experiments on the properties of the 2016 Competition on Runtime Verification have shown promising results. As we have observed, a majority of the tested properties benefit from a speed boost of 2$\times$ and more through the careful application of the patterns presented in this paper. These encouraging numbers warrant further development of this line of research in multiple directions. First, 
modifications to the existing patterns could be developed to create longer-lived threads and reduce the overhead incurred by their creation and destruction. Second, one could relax the way in which properties are evaluated, and tolerate that finite prefixes of an output trace be different from the exact result. This notion of ``eventual consistency'' could allow multiple slices of an input to be evaluated without the need for lock-step synchronization. Finally, special care should be given to the way queries are expressed; when several equivalent representations of the same computation exist, one should favor those that lend themselves to parallelization.


\bibliographystyle{abbrv}



\end{document}